\documentclass[conference]{IEEEtran}
\IEEEoverridecommandlockouts
\usepackage{cite}
\usepackage{amsmath,amssymb,amsfonts}
\usepackage{graphicx}
\usepackage{textcomp}
\usepackage{xcolor}
\usepackage{subfigure}
\usepackage{tablefootnote}
\usepackage{booktabs}
\usepackage{makecell}
\usepackage{stfloats}
\usepackage{pifont}
\usepackage{algorithm}  
\usepackage[normalem]{ulem}  
\usepackage{algpseudocode}  
\usepackage{amsmath}  
\usepackage{multirow}
\usepackage{fix-cm}
\usepackage{csquotes}
\usepackage{todonotes}

\usepackage{siunitx}
\usepackage{tikz}

\usetikzlibrary{shapes.geometric, arrows}

\tikzstyle{startstop} = [rectangle, rounded corners, 
minimum width=3cm, 
minimum height=1cm,
text centered, 
text width=3.5cm, 
draw=black, 
fill=gray!30]
\tikzstyle{process} = [rectangle, 
minimum width=3cm, 
minimum height=1cm, 
text centered, 
text width=3cm, 
draw=black, 
fill=gray!30]
\tikzstyle{optional} = [rectangle, dashed, 
minimum width=3cm, 
minimum height=1cm, 
text centered, 
text width=3cm,
draw=black,
fill=white!30,
]
\tikzset{decision/.style={
    draw=black,
    fill=gray!30,
    diamond,
    text width=1.5cm, 
    align=center,
    aspect=1,
    inner sep=0pt, 
    minimum width=2cm, 
    minimum height=1cm 
}}
\tikzstyle{arrow} = [thick,->,>=stealth,line width=1.0mm]

\def\BibTeX{{\rm B\kern-.05em{\sc i\kern-.025em b}\kern-.08em
    T\kern-.1667em\lower.7ex\hbox{E}\kern-.125emX}}

\begin{document}

\title{Large Language Models (LLMs) for \\Electronic Design Automation (EDA) \vspace{-0.4cm}\\
\vspace{5pt}
\textit{\fontsize{12pt}{12pt}\selectfont Special Session Paper \vspace{-0.4cm}}
}

\author{
\IEEEauthorblockN{
\fontsize{10pt}{10pt}\selectfont
Kangwei Xu\textsuperscript{1}, 
Denis Schwachhofer\textsuperscript{2}, 
Jason Blocklove\textsuperscript{3},
Ilia Polian\textsuperscript{2}, 
Peter Domanski\textsuperscript{2}, 
Dirk Pflüger\textsuperscript{2},\\
Siddharth Garg\textsuperscript{3},
Ramesh Karri\textsuperscript{3},
Ozgur Sinanoglu\textsuperscript{3}, 
Johann Knechtel\textsuperscript{3},
Zhuorui Zhao\textsuperscript{1}, 
Ulf Schlichtmann\textsuperscript{1},
Bing Li\textsuperscript{4}
}
\vspace{0.2cm}\IEEEauthorblockA{\textsuperscript{1}\textit{Technical University of Munich~}\hspace{5pt}\textsuperscript{2}\textit{University of Stuttgart~}\hspace{5pt}\textsuperscript{3}\textit{New York University~}\hspace{5pt}
\textsuperscript{4}\textit{University of Siegen}}\\
\vspace{-1.25cm}}

\maketitle

\vskip -20pt

\begin{abstract}
With the growing complexity of modern integrated circuits, hardware engineers are required to devote more effort to the full design-to-manufacturing workflow. This workflow involves numerous iterations, making it both labor-intensive and error-prone. Therefore, there is an urgent demand for more efficient Electronic Design Automation (EDA) solutions to accelerate hardware development. Recently, large language models (LLMs) have shown remarkable advancements in contextual comprehension, logical reasoning, and generative capabilities. Since hardware designs and intermediate scripts can be represented as text, integrating LLM for EDA offers a promising opportunity to simplify and even automate the entire workflow.
Accordingly, this paper provides a comprehensive overview of incorporating LLMs into EDA, with emphasis on their capabilities, limitations, and future opportunities.
Three case studies, along with their outlook, are introduced to demonstrate the capabilities of LLMs in hardware design, testing, and optimization.
Finally, future directions and challenges are highlighted to further explore the potential of LLMs in shaping the next-generation EDA, providing valuable insights for researchers interested in leveraging advanced AI technologies for EDA.
\end{abstract}


\section{Introduction}
Electronic Design Automation (EDA) spans the entire workflow from logic design to manufacturing, which plays a key role in improving hardware performance 
and shortening development cycles. The advent of artificial general intelligence (AGI) brings revolutionary changes to reshape the future of EDA \cite{b0}. By leveraging techniques such as large language models (LLMs), patterns can be learned, and insights can be inspired from a large amount of historical data, providing more efficient and intelligent solutions for the EDA workflow. Since circuits can be represented by hardware description languages (HDL)
and scripts are inherently text-based, these characteristics 
align well with LLM capabilities, creating new opportunities to alleviate labor-intensive tasks in the EDA workflow~\cite{b1, b2, b3, b3.1, b3.2}.

\begin{figure}[]
\centering	\includegraphics[width=1\linewidth]{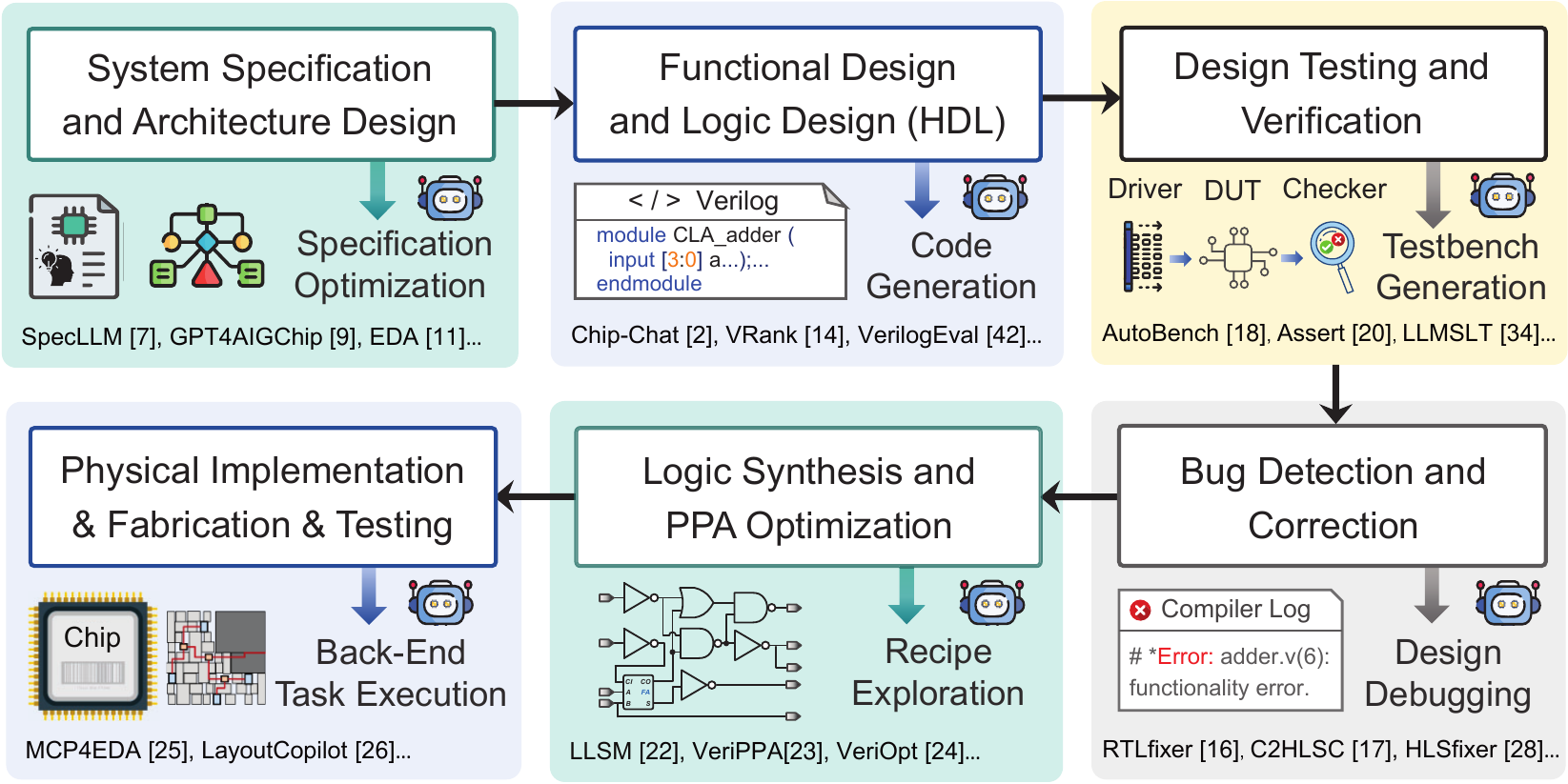}
\vspace{-0.6cm}
\caption{Typical chip design flow and potential LLM applications.}
\label{fig:eda}
\vspace{-0.65cm}
\end{figure}

Recent studies show that applying LLMs to EDA has great potential to transform the interaction between designers and tools from passive assistance to active agents capable of autonomously performing specific tasks~\cite{b4.2, b5, b6}. Fig.~\ref{fig:eda} illustrates their applications across the chip design flow, including tasks such as specification optimization and hardware generation, \textit{etc}. Standing at the frontier of next-generation EDA, the emergence of LLMs is regarded as a transformative progress that can enhance hardware development efficiency and foster innovation, with the potential to fundamentally redefine the design, verification, and manufacturing of electronic systems.

While LLMs bring many benefits to the EDA workflow, they would be widely adopted only if they can achieve groundbreaking advancements to address the complexities of hardware design. Current approaches still fall short of an integrated design synthesis, often neglecting the interdependencies between logic and its physical implementation. Achieving this goal requires enabling these multi-modalities to co-exist and interact, ensuring smooth transitions between various design stages, which remains highly challenging. In this paper, a comprehensive overview of how LLMs are shaping the next-generation EDA is provided, offering clear insights to advance AI integration and evaluating whether this integration is an actual breakthrough or an overestimated future.


The organization of this paper is as follows. Section II presents a comprehensive discussion of the state-of-the-art application of LLMs in EDA.
Section III explores the use of LLMs for hardware design and testing.
Section IV provides the evolution of LLM-aided hardware design approaches, from finetuned models to structured prompting frameworks.
Section V uses LLMs with intelligent prompts to automatically generate System-Level Test programs.
Section VI explores the future prospects and challenges of LLM applications in EDA, underscoring both their limitations and opportunities in this fast-developing domain. Section VII concludes the paper.

\section{State of the Art of LLM Applications in EDA}
Hardware design typically begins with design specifications, which are then written into hardware description languages (HDL) by experienced engineers. In modern hardware development, high-level synthesis (HLS) translates C/C++/SystemC into RTL, while hardware construction languages such as Chisel generate parameterized RTL through high-level abstractions~\cite{b7}. Both approaches significantly improve design flexibility and efficiency. Once RTL is generated, static analysis and verification ensure correctness, followed by logic synthesis to a Power, Performance, and Area (PPA)-optimized gate-level netlist and physical design for a manufacturable layout. Despite these advances, the process remains error-prone, time-consuming, and heavily dependent on human effort~\cite{b8}.

Recently, LLMs demonstrate remarkable capabilities in context understanding and logical reasoning, making it possible to assist engineers across EDA tasks from high-level design specification to low-level physical implementation~\cite{b4.2, b5, b6}.

In front-end hardware design, Chip-Chat~\cite{b1} uses GPT-4 to complete a full HDL tapeout for an 8-bit microprocessor, enhanced by hierarchical prompting for complex designs~\cite{b2}. GPT4AIGChip~\cite{b5} automates AI accelerator development by decoupling hardware modules and incorporating LLM-friendly templates for iterative optimization. VRank and VFocus~\cite{b9,vfocus} exploits the probabilistic nature of LLMs to generate multiple Verilog candidates, cluster them by simulation outputs, rank them by consistency, and select the best design.

The frontier of LLM-driven hardware design is further advanced in HLS. C2HLSC~\cite{b10} uses LLMs to iteratively convert C programs into HLS-compatible versions with tool-guided feedback and hierarchical decomposition. HLS-Repair~\cite{b19.1} applies LLMs with Retrieval-Augmented Generation and optimization strategies to convert C code into optimized HLS-C, reducing human effort while improving synthesis quality. 

Hardware testing validates the functional behavior of generated hardware. AutoBench and CorrectBench~\cite{b13,correctbench} use LLMs to implement a hybrid test platform and self-testing system, which is later improved through a self-correction loop. 
System-Level Test (SLT) detects defects missed by earlier tests~\cite{polian2020} but still relies heavily on manual effort~\cite{chen2018}. To mitigate this, LLMs with Structural Chain of Thought prompting~\cite{li2023scot} are used to automatically generate C code that maximizes power consumption in a superscalar, out-of-order RISC-V processor on an FPGA, enabling defect detection under high activity or temperature.

Hardware verification ensures design correctness through assertion checking. AssertLLM~\cite{b14} generates assertions from complete specifications by extracting structures and mapping signals. AutoSVA~\cite{b15} uses an iterative framework with formal verification feedback to refine LLM-generated assertions.

Through logic synthesis, the RTL design is then transformed into an optimized gate-level representation. LLSM~\cite{b16} introduces an LLM-enhanced logic synthesis model capable of directly extracting information from RTL code, integrating Chain-of-Thought prompting, text–circuit hybrid embeddings, and an AIG-tailored acceleration library to improve synthesis efficiency and quality. 

Physical implementation transforms a synthesized design into a manufacturable layout. 
MCP4EDA~\cite{b17} provides an LLM-driven framework that controls the RTL-to-GDSII flow via natural language and uses PPA metrics for iterative script optimization.
LayoutCopilot~\cite{b18} proposes an LLM-powered multi-agent framework to convert natural language into executable script commands for interactive analog layout design.


\section{LLM-Assisted Design and Testbench Generation}

\begin{figure}[]
\centering	\includegraphics[width=1\linewidth]{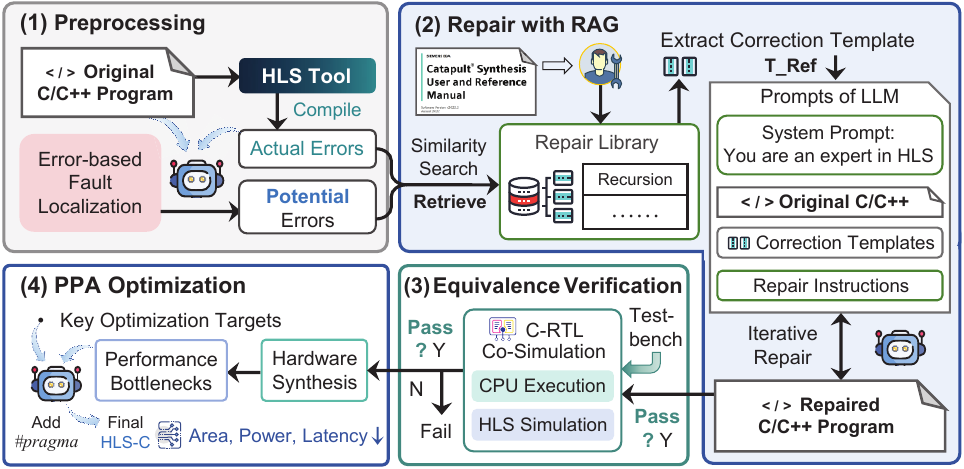}
\vspace{-0.7cm}
	\caption{Automated C/C++ program repair  with LLMs for HLS.}
	\label{fig:rag}
\vspace{-0.35cm}
\end{figure}

\begin{figure}[]
\centering	\includegraphics[width=1\linewidth]{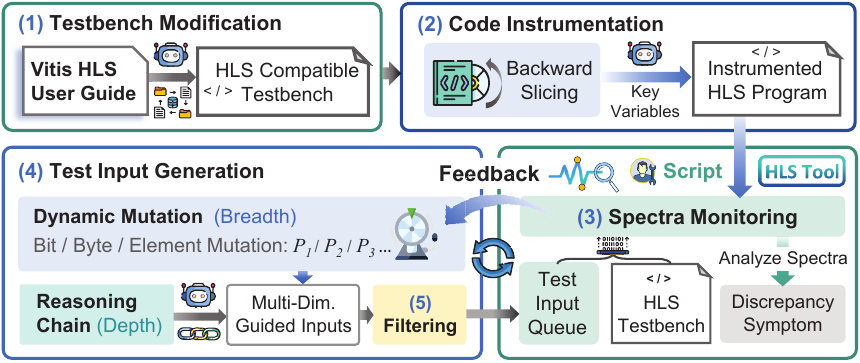}
\vspace{-0.7cm}
	\caption{Efficient testing of behavioral discrepancies with LLMs for HLS. }
	\label{fig:test}
\vspace{-0.65cm}
\end{figure}

High-Level Synthesis (HLS) takes computer programming languages such as C/C++ as input and automatically outputs HDLs like Verilog. However, converting regular C/C++ programs to HLS-compatible versions (HLS-C), which can be synthesized by HLS tools, often requires significant human effort~\cite{b19}. For instance, dynamic memory allocation should be rewritten by engineers, as hardware cannot handle unbounded data structures. Recently, as shown in Fig.~\ref{fig:rag}, an LLM-aided program repair framework for HLS~\cite{b19.1} is proposed to fix such incompatibilities in C/C++ code, which includes four stages: 

\textit{(1) Preprocessing:} The original C/C++ program is compiled by an HLS tool, and some actual errors are returned. Since the HLS compiler may not be able to detect all errors in one go, an LLM is used for other potential error detection.

\textit{(2) Repair with RAG:} Retriever-Augmented Generation (RAG) enhances LLM capability by integrating expert knowledge via a retriever. Incorporating retrieved correction templates from the external library in the LLM's prompts effectively guides the LLM towards accurate C program repairs.

\textit{(3) Equivalence Verification:} After synthesizing the repaired C program into the corresponding RTL code, a C-RTL co-simulation is performed to verify functional equivalence.

\textit{(4) PPA Optimization:} Repaired HLS-C programs are collected for further optimization, where the LLM optimizes code segments with performance bottlenecks by adjusting pragmas.

While the above framework converts regular C/C++ programs to HLS-compatible versions, these HLS-C programs and the resulting FPGA-deployed circuits may still exhibit behavioral discrepancies compared with the original CPU execution. These discrepancies arise from assumptions made during the repair and synthesis processes. For example, customized bit widths in FPGA deployment may cause overflows, while pipeline parallelism (\texttt{\#pragma HLS pipeline}) could introduce results that deviate from sequential CPU execution due to data dependencies or feedback paths. To address this challenge, an LLM-aided framework for testing behavioral discrepancy in HLS is proposed~\cite{b20}. As shown in Fig.~\ref{fig:test}, the original C/C++ testbench is first adapted into an HLS-compatible testbench with the aid of the LLM by removing unsupported C/C++ syntax (1), so that it can be successfully compiled by the HLS tool. Then, a backward slicing technique is applied to identify key variables (2), which are then instrumented by the LLM to enable spectra monitoring (3). The collected spectra are fed into the test input generation that combines dynamic mutation with an LLM-based reasoning chain (4). Finally, a redundancy filtering method is employed to skip repeated hardware simulations (5). 

These two LLM-aided frameworks target program repair and behavioral discrepancy testing for HLS, potentially advancing the efficiency of design, debugging, and optimization. Looking ahead, the goal is accurate, automated translation from high-level descriptions to optimized RTL, rivaling or surpassing the performance of expert-crafted RTL designs, ultimately enabling a seamless end-to-end design flow.

\section{Large Language Models for Hardware Design: Opportunities and Challenges}
The first exploration of using LLMs for hardware design took place in 2020 with a finetuned GPT-2 model known as DAVE~\cite{pearce2020dave}.
DAVE was trained using simple textbook-style Verilog questions, generally considered to be novice-level problems, and was ultimately very successful at solving similarly defined and simple problems, but significantly struggled with more complex designs or more open-ended tasks.
To address the shortcomings of this model, a new family of models called VeriGen was proposed~\cite{thakur2023verigen}.
This was a set of five finetuned models, with the most capable of the models being finetuned from the CodeGen family of models, created using both the content from several textbooks as well as a significant amount of open-source Verilog code from GitHub.
The CodeGen models outperformed  ChatGPT-3.5 and performed similarly well to GPT-4, both from OpenAI, at a fraction of the model size.
Following the release of these models, numerous more LLMs finetuned for Verilog were created, each attempting to improve performance or reduce computational cost in their own way.
These models include RTLCoder~\cite{rtlcoder}, VerilogEval~\cite{verilogeval}, and CodeV~\cite{codev}.

Each of these finetuned models, however, is more traditional ``autocompletion-style'' LLMs, and with the release of ChatGPT in 2022 the development focus heavily shifted towards more ``conversational'' LLM use with instruction-tuned models.
Rather than a model merely trying to guess the next set of tokens the user might want, these models wait until a complete thought has been presented and then aim to respond completely.
In Chip-Chat~\cite{b1}, the authors used ChatGPT-4, a general-knowledge conversational model from OpenAI, to help design and fully generate the Verilog for a novel accumulator-based ISA designed to fit in a heavily space-constrained tapeout through TinyTapeout.
ChatGPT was used to help create the specification of the ISA as well as generate 100\% of the Verilog and a Python assembler for testing.
This was ultimately taped out and verified to fully work as anticipated, making this the first design taped out with the hardware fully written by AI.

Chip-Chat, however, was generally unstructured and relied on an experienced hardware designer to guide the development.
These issues led to the development of a more structured framework for using LLMs to generate Verilog, as well as using the LLM to generate testbenches to verify that the Verilog worked as expected~\cite{b6}.
To accomplish this, a strict conversational prompting method was adopted, first asking the LLM to generate a design, then requesting a testbench for the design, and then using Icarus Verilog to verify the functionality using the testbench.
\begin{figure}[t]
    \centering
	\includegraphics[width=\linewidth]{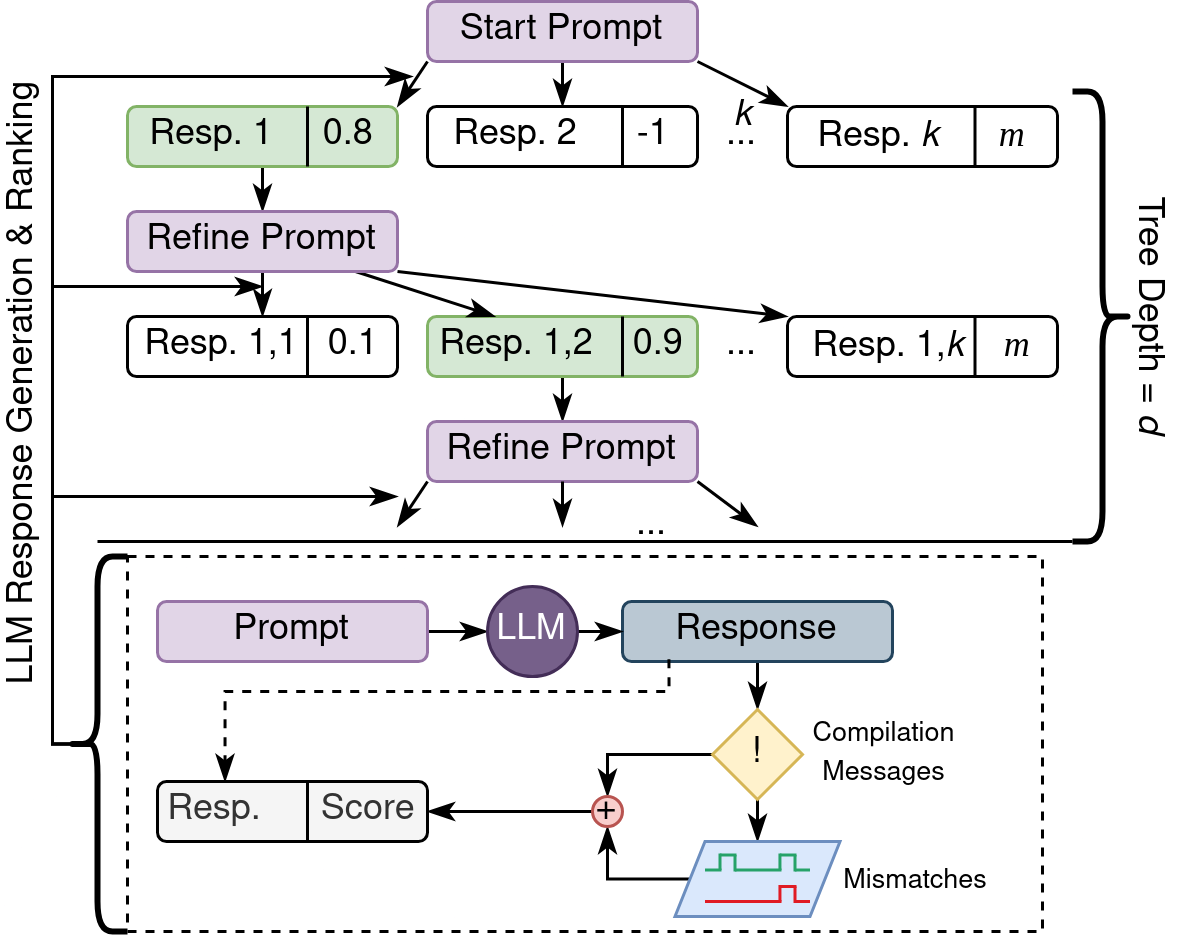}
    \vspace{-0.55cm}
    \caption{AutoChip design framework with tree-search~\cite{b4.2}.}\label{fig:autochip-tree}
    \vspace{-0.7cm}
\end{figure}
To minimize the potentially nebulous human feedback, the resulting output from compiling and simulating the design was fed back into the LLM to fix any issues that occurred, with human feedback only given when a mistake failed to be corrected numerous times.
This framework was used with both ChatGPT-3.5 and -4, which were the most advanced models available at the time, and the set of 8 benchmark designs to be generated was kept rather simple.
This technique showed promise, with half of the tests run on ChatGPT-4 requiring no human feedback at all, but significant issues were identified with the generated testbenches lacking acceptable test coverage.

Following the partial success of the structured feedback-driven design flow, it became clear that the next step in using LLMs to design hardware was to fully remove the human from the loop and automate the whole process.
This resulted in AutoChip~\cite{b4.2}, a fully-automated Verilog design tool leveraging LLMs and open-source EDA tools to generate complete designs.
The AutoChip framework is shown in Figure~\ref{fig:autochip-tree}.

Given the lack of testbench quality, AutoChip now requires a quality testbench that provides feedback on design errors as input from the user.
To evaluate the capabilities of AutoChip as a framework, the VerilogEval benchmark set was utilized, which contains both prompts and well-formatted testbenches for use such as this.
Beyond just iterating over a design to improve it, AutoChip also introduced the use of a tree-search functionality, where $k$ candidate responses were gathered and evaluated using the EDA tools and testbenches, the responses were ranked, and the best response was used for the feedback.
The ranking was performed as a percentage of passing test cases in a design's testbench.

The tree-search design was a result of observations that if an LLM was substantially incorrect early in the design, it was less likely to properly recover than one that was initially closer.
These designs were iterated up to a tree depth of $d$, and the most successful response was then given as an output.

AutoChip was evaluated using four state-of-the-art commercially available LLMs, and it was found that the most capable model at the time, GPT-4o, was the only one to significantly benefit from the use of feedback over generating a substantial number of candidate responses.
This likely points to a lack of LLM training on the meanings and fixes for EDA tool error messages, with only the most capable models being able to determine how to properly leverage the feedback.

Similar automation techniques and designs using this automation have been shown following AutoChip.
In~\cite{b2}, the CL-Verilog model is proposed, a finetuned Code Llama model for generating Verilog, and with it, a framework is given which hierarchically builds designs using smaller modules as building blocks to enable the generation of more realistic designs for practical applications, such as cryptographic accelerators.

\section{When LLMs Test Chips: Language Models for the Automation of System-Level Test Programs}
System-Level Test (SLT) has become a crucial part of the manufacturing process for integrated circuits over the last decade.
In SLT, the Device Under Test (DUT) is placed in an environment that mimics its real-world application~\cite{chen2018}.
For example, a System-on-Chip (SoC) planned for use in smartphones is put onto a board that resembles an actual smartphone.
Off-the-shelf software is executed on the DUT to observe errors or crashes, with test suites manually composed by test engineers based on experience and field returns.
To continue the example: We then would boot an Android and, if successful, continue using different apps, take a call, or stream a video, and more.
If no unexpected errors or crashes occur in the end, the DUT is shipped to the customer.
For SLT, we assume that the DUT is a black box, since external entities, for example, integrators, may also engage with it.

SLT aims to detect defects not caught by structural tests by exercising additional paths and transactions.
However, some defects require specific non-functional conditions to be fulfilled, e.g., power consumption or temperature, to be detected~\cite{polian2020}.
Such defects are called marginal defects.
It is highly challenging to manually write high-level code that can control these non-functional properties, especially without extensive knowledge about the DUT.

Automatically generating code in a high-level language such as C is highly challenging.
There are methods that can reliably generate assembly code controlling non-functional properties~\cite{schwachhofer2023fuzz,schwachhofer2024gp}; however, one could make the argument that such snippets will most likely not occur in real-world software, whereas one aspect of SLT is the approximation of the end-user environment.
LLMs alleviate this challenge in two ways: First, they simplify C code generation; second, their training on real-world applications makes them more likely to generate high-quality code.

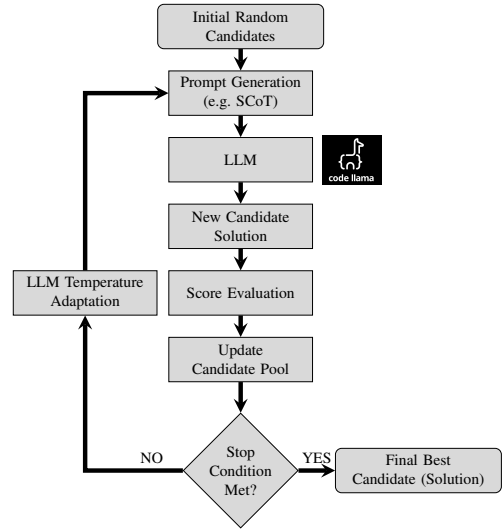
\begin{figure}[t]
    \centering
    \resizebox{0.75\linewidth}{!}{
        \begin{tikzpicture}[node distance=1.5cm, auto]
            \node (start) [startstop] {Initial Random Candidates};
            \node (n1) [process, below of=start] {Prompt Generation (e.g. SCoT)};
            \node (llm) [process, below of=n1] {LLM};
            \node (code_llama) [inner sep=0pt, right of=llm, xshift=1cm] {\includegraphics[width=0.15\linewidth]{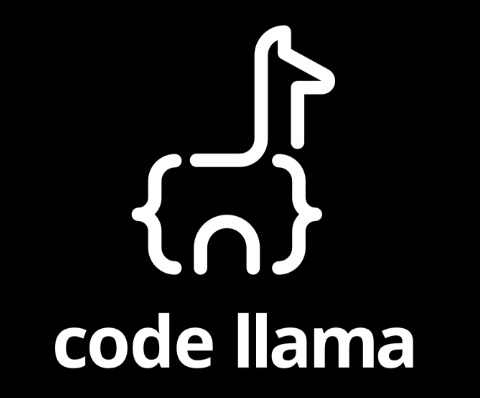}};
            \node (n2) [process, below of=llm] {New Candidate Solution};
            \node (n3) [process, below of=n2] {Score Evaluation};
            \node (n4) [process, below of=n3] {Update Candidate Pool};
            \node (n5) [process, left of=n3, xshift=-2cm] {LLM Temperature Adaptation};
            \node (d1) [decision, below of=n4, yshift=-1cm] {Stop Condition Met?};
            \node (end) [startstop, right of=d1, xshift=2.5cm] {Final Best Candidate (Solution)};
    
            \draw [arrow] (start) -- (n1);
            \draw [arrow] (n1) -- (llm);
            \draw [arrow] (llm) -- (n2);
            \draw [arrow] (n2) -- (n3);
            \draw [arrow] (n3) -- (n4);
            \draw [arrow] (n4) -- (d1.north);
            \draw [arrow] (d1.west) -| node[anchor=south, xshift=1.5cm]{NO} (n5.south);
            \draw [arrow] (n5.north) |- (n1.west);
            \draw [arrow] (d1.east) -- node[anchor=south]{YES} (end);            
        \end{tikzpicture}
    }
    \vspace{-0.25cm}
    \caption{Flow of the optimization loop.}\label{fig:sltopt}
    \vspace{-0.7cm}
\end{figure}

We have developed an approach, see Fig.~\ref{fig:sltopt}, to generate C code with the goal of maximizing the power consumption of a superscalar, out-of-order RISC-V processor, called BOOM~\cite{zhao2020}, running on an FPGA~\cite{schwachhofer2024llm}.
We use the FPGA to measure the power consumption of the generated C code.
More about the measurement setup can be found in~\cite{schwachhofer2023fuzz}; it should be noted that some improvements have been applied.
Initially, we provide a handwritten set of programs as examples.
The prompt is then generated using \(n\) randomly picked examples from the candidate pool and is passed to the LLM.
The response is evaluated, either using a microarchitectural simulator or power measurements.
The score is set to zero if the code does not compile or causes an unwanted exception.
Based on the evaluation results, the new snippet is either added to the candidate pool or discarded.
We then check if any stop condition is fulfilled, for example, the number of snippets, time, or the user stopping the process manually.
If it is not stopped, the temperature of the LLM is adapted to the result, and the loop starts again.

Our approach uses \enquote{Structural Chain of Thought} (SCoT)~\cite{li2023scot} prompting to increase the quality of the output.
The idea behind SCoT is that we first let the LLM generate pseudocode that it thinks fulfills the requirements we state.
In a second prompt, we use the generated pseudocode to ask it for proper C code.
We also hint that there might be errors in the pseudocode.
The examples we supply also contain the power consumption of the snippet to give the LLM a better idea which of the examples is better and which to avoid.

For our LLM, we used Code Llama with 34 billion parameters in the \enquote{Instruct} flavor that has been further fine-tuned by an external company, incorporating 80k question-answer pairs and an additional 1.5 billion tokens.
Compared to the off-the-shelf model, it performs significantly better.

We have implemented a temperature adaptation mechanism to guide the LLM further towards better snippets.
Lower temperature allows the LLM to focus more on improving the examples from the candidate pool (exploitation), while a higher temperature allows it to generate more diverse code snippets (exploration).
The idea is borrowed from simulated annealing.
The adaptation follows a dynamic schedule that depends on the score of the generated snippet as well as its Levenshtein distance to the other snippets in the pool.
The Levenshtein distance is introduced to force the pool to be more diverse, because otherwise the LLM will converge towards very similar snippets and become stuck in a local optimum.
More details are presented in~\cite{schwachhofer2024llm}.

Our optimization loop ran for 24 hours and produced 2021 snippets.
The best snippet in this run consumes \qty{5.042}{\watt}.
However, we have run genetic programming (GP) for 39 hours, finding a snippet that consumes \qty{5.682}{\watt}.
The discrepancy in runtime comes from the fact that after 24 hours, GP still found better results, whereas multiple previous experiments showed that for the LLM-based approach, significant changes rarely, if at all, happen.
We assume that the difference, \qty{0.640}{\watt}, comes from multiple factors, including the compiler itself, the flags passed to the compiler, the LLM itself, or rather, its fine-tuning, and finally, the language used.
Despite existing limitations, the LLM-based approach is promising, especially when considering that the GP snippet has no real-world equivalent.
Nonetheless, future work should explore these limitations to understand if there are more improvements possible.

\section{Future Directions and Challenges}

While LLMs in EDA have shown significant progress, the field is still in its early stages, and the integration of LLMs into EDA faces many challenges but also opportunities. 

\begin{figure}[t]
\centering	\includegraphics[width=1\linewidth]{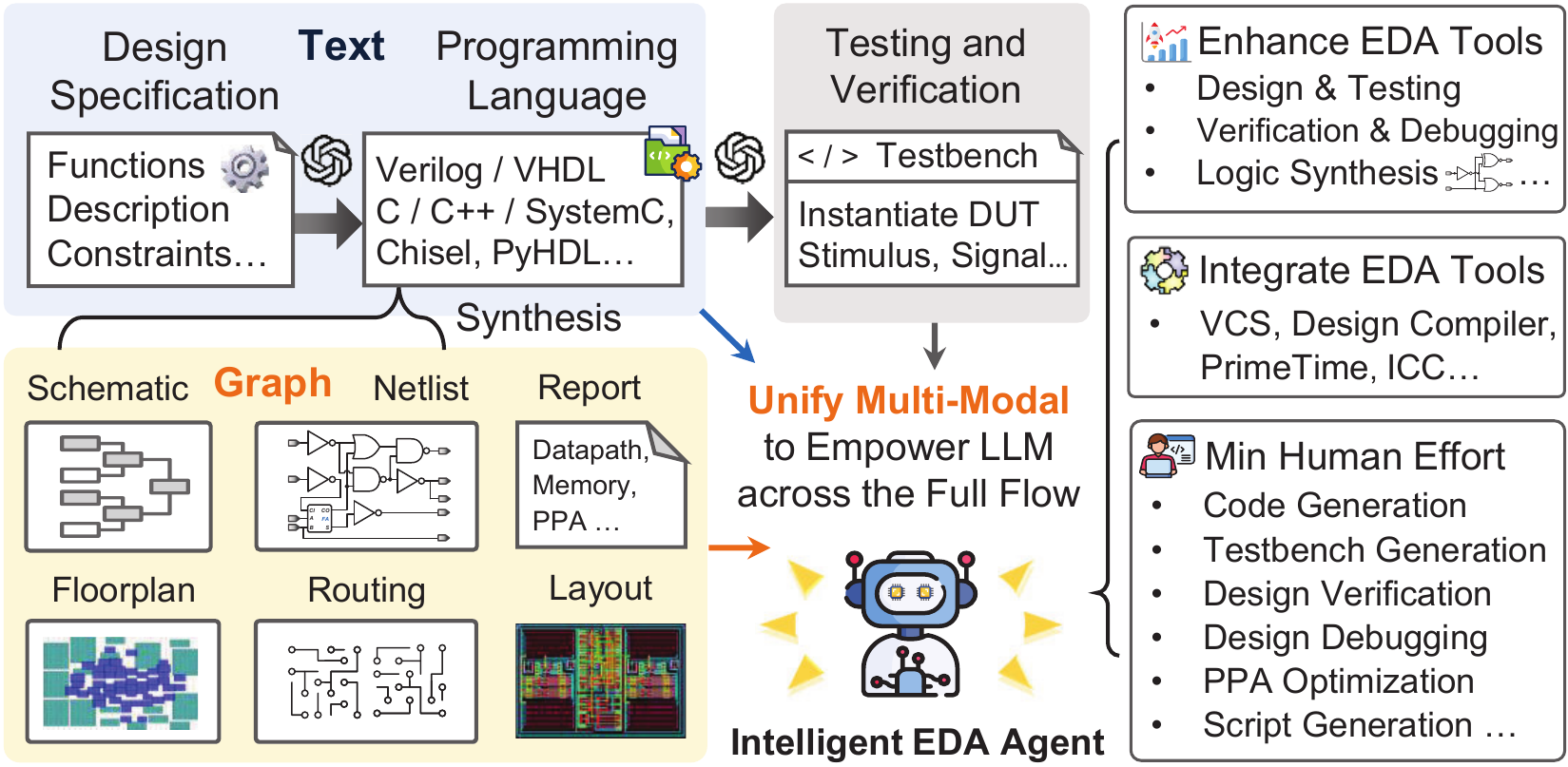}
\vspace{-0.55cm}
	\caption{LLM-aided EDA Agent: Unifying Multi-Modal Across the Full Flow}
	\label{fig:agent}
\vspace{-0.6cm}
\end{figure}

As shown in Fig.~\ref{fig:agent}, we envision an LLM-powered intelligent agent designed not merely to augment but to fundamentally reshape the EDA workflow. Unlike conventional AI4EDA approaches that overlook the complexities of circuit design across different stages, this agent integrates natural language specifications, HDL designs, and multi-modal data, such as schematics, netlists, and physical layouts, into a unified representation. This enables a deeper understanding of design intent and seamless integration of various EDA tools. By capturing the interdependencies between logic and physical implementation, the agent supports comprehensive synthesis, full automation, and generalizable solutions for the entire EDA workflow. Here, we propose several directions that can be further explored to advance the development of LLMs in EDA.

$\circ$~\textit{\textbf{Bridging Semantic Gaps:}} Despite the multi-modal representations across circuit design stages, the underlying functions remain consistent. The specification-to-silicon process requires accurately mapping functional behaviors across natural language specifications, high-level descriptions (e.g., Python/C/C++), RTL designs, and physical implementations. In traditional workflows, this process is both time-consuming and error-prone, due to the persistent semantic gap between high-level abstractions and hardware implementations. Bridging this gap remains a long-term challenge in the EDA community. Developing advanced feature extraction and alignment techniques, enhanced by LLM-driven semantic analysis, could bridge the semantic gaps, enabling more integrated, comprehensive, and efficient hardware design methodologies.

$\circ$~\textit{\textbf{Toward Expert-Level HLS:}} While HLS enables efficient hardware design in C/C++ at a higher abstraction level, its conventional compilation pipelines often fall short of the Quality of Results (QoR) achieved by the hand-crafted HDL. This performance gap stems from imprecise hardware semantics in C/C++, limited co-optimization between logic design and physical implementation, and the expertise gap between HLS compilers and experienced hardware engineers. We envision an LLM-enhanced HLS agent that internalizes expert hardware design heuristics by learning from optimal paired HLS–HDL codebases, annotated hardware directives, synthesis reports, and post-layout feedback. By integrating closed-loop QoR prediction and iterative refinement, such an LLM-guided HLS agent could progressively converge toward, and in some cases surpass, expert-crafted HDL, bridging the performance gap between next-generation HLS and experienced HDL design.

$\circ$~\textit{\textbf{High-Level Guided RTL Debugging:}} LLM-generated HDL designs are still prone to functional errors due to limited high-quality HDL datasets and the fine-grained nature of hardware constraints. Manual debugging is time-consuming, relying heavily on simulation and exhaustive waveform inspection. In contrast, LLMs show high accuracy in producing untimed behavioral models in languages like Python or C/C++. Leveraging this strength, an LLM can generate functionally equivalent high-level descriptions from natural languages, enabling cross-level comparison with RTL simulations. Such validation uses reliable high-level execution as a reference to effectively compensate for error-prone HDL generation.

$\circ$~\textit{\textbf{Intelligent Kernel Extraction for  Accelerator Generation:}} In traditional HLS workflows, identifying and extracting compute-intensive kernels for accelerator generation is an expert-driven and time-intensive process. Generating high-performance accelerators becomes especially challenging for designs with complex control flow or irregular memory access. Moreover, inefficient CPU–accelerator data transfer can negate the performance gains of hardware acceleration. An LLM-driven agent could integrate kernel detection, PPA optimization, and iterative refinement into a closed-loop process, accelerating development while improving design quality.

$\circ$~\textit{\textbf{Privacy and Security:}} Engineers may rely on cloud-based LLMs to automatically generate hardware solutions, which could expose sensitive data to privacy breaches or intellectual property theft~\cite{zeng24_sec,zeng25_leak}. Moreover, malicious code or hardware Trojans may be inserted into the generated hardware designs via the cloud platform~\cite{mankali25_HT}. Deploying LLMs on local servers, combined with privacy-aware interaction with cloud-based models~\cite{zeng25_unlearn}, could help to protect sensitive data.

$\circ$~\textit{\textbf{Seamless Integration of EDA Tools:}} Modern EDA workflow spans from design specifications to physical implementations, requiring multiple modalities to capture full design intent. Text conveys logic, while visuals reveal structural complexity, enabling LLMs to interpret complex designs. However, inconsistent formats, interfaces, and workflows continue to hinder seamless EDA tool integration. Unifying multi-modal representations can align high-level logic with low-level implementation. Real-time feedback and comprehensive results could also be integrated to iteratively optimize the generated hardware. Such an agent would not merely enhance existing EDA tools but fundamentally revolutionize them, opening a new era of intelligent, fully automated hardware design.

\section{Conclusion}
Integrating Large Language Models (LLMs) into the EDA workflow opens new possibilities for exploring a wider range of automatic hardware design processes, with the potential to redefine the paradigm of circuit design. By capturing the complex characteristics of large-scale circuits, next-generation EDA tools powered by LLMs can enable more accurate and innovative design strategies, thereby not only reducing hardware development costs but also accelerating time-to-market. Despite challenges such as LLM hallucinations, the inherent complexity of hardware design, and data privacy concerns, the application of LLMs in EDA still represents a transformative opportunity, paving the way toward more efficient and intelligent hardware development and ultimately reshaping the future of EDA.

\section*{Acknowledgements}
This work was supported in part by Advantest as part of the Graduate School \enquote{Intelligent Methods for Test and Reliability} (GS-IMTR) at the University of Stuttgart and by the Deutsche Forschungsgemeinschaft (DFG, German Research Foundation) – Project-ID 497488621.

\end{document}